\documentclass[12pt]{iopart}
\usepackage{tabularx} 
\usepackage{amsmath}  
\usepackage{graphicx} 
\usepackage[margin=1in,letterpaper]{geometry} 
\usepackage{cite} 
\usepackage[final]{hyperref} 
\usepackage{breqn}
\usepackage{float}
\usepackage{comment}
\hypersetup{
	colorlinks=true,       
	linkcolor=blue,        
	citecolor=blue,        
	filecolor=magenta,     
	urlcolor=blue         
}

\usepackage{color}

\usepackage{amsmath}
\usepackage{slashed}
\usepackage{amssymb}
\usepackage{subcaption}
\usepackage{cases}
\usepackage{xcolor}
\numberwithin{equation}{section}

\begin{document}

\title{Detection of high-frequency gravitational waves using high-energy pulsed lasers}

\author{Georgios Vacalis}
\address{Department of Physics, University of Oxford, Oxford OX1 3PU, UK}
\author{Giacomo Marocco}
\address{Department of Physics, University of Oxford, Oxford OX1 3PU, UK}
\address{Lawrence Berkeley National Laboratory, 1 Cyclotron Rd, Berkeley CA 94720, USA}
\author{James Bamber}
\address{Department of Physics, University of Oxford, Oxford OX1 3PU, UK}
\author{Robert Bingham}
\address{Rutherford Appleton Laboratory, Chilton, Didcot, Oxon OX11 OQX, UK}
\address{Department of Physics, University of Strathclyde, Glasgow G4 0NG, UK}
\author{Gianluca Gregori}
\address{Department of Physics, University of Oxford, Oxford OX1 3PU, UK}

\ead{georgios.vacalis@physics.ox.ac.uk}
\vspace{10pt}
\begin{indented}
\item[]May 2023
\end{indented}

\begin{abstract}
We propose a new method for detecting high-frequency gravitational waves (GWs) using high-energy pulsed lasers. Through the inverse Gertsenshtein effect, the interaction between a GW and the laser beam results in the creation of an electromagnetic signal. The latter can be detected using single-photon counting techniques. We compute the minimal strain of a detectable GW which only depends on the laser parameters. We find that a resonance occurs in this process when the frequency of the GW is twice the frequency of the laser. With this method, the frequency range $10^{13}-10^{19} $ Hz is explored non-continuously for strains $h \gtrsim 10^{-20}$ for current laser systems and can be extended to $h \gtrsim 10^{-26}$ with future generation facilities. 
\end{abstract}
\vspace{3cm}

\newpage

\section{Introduction}
In 2015, the LIGO and the Virgo collaborations made the first observation of a gravitational wave (GW) signal \cite{Ligo} almost a century after the existence of GWs was theoretically predicted by Einstein. The signal was generated by a binary black hole (BH) merger and covered a range of frequencies from a few tenths up to a few hundred Hz. This discovery opened a new path for observing physical phenomena that happened early in the history of the Universe, as well as more recent ones. Theoretically, GWs can be emitted at any frequency. Current ground-based interferometers can detect GWs in the range 10 Hz - 10 kHz \cite{Ligo2, Virgo, GEO, KAGRA}, whereas space-based interferometers like LISA are expected to cover the range 0.1 mHz - 1 Hz \cite{LISA}. Frequencies above 10 kHz have not been explored because detectors such as LISA have not been targeted for this range. Moreover, the sensitivity of LIGO and Virgo decreases in the same range due to the cavity response which becomes worse as the GW frequency increases. Nonetheless, they are of particular interest \cite{potentialvhf}. For example, observing high-frequency GWs can help with the detection of extra dimensions \cite{extradim}. On the other hand, several sources that produce high-frequency GWs correspond to events in the early Universe, see Refs.~\cite{cosmoback,Challenges} for reviews. As an example of an event we refer to the evaporation of primordial BHs which emits GWs whose spectrum peaks for frequencies beyond 1 THz \cite{pbh, pbh2}. Hypothetical dark matter candidates consisting of gravitationally bound atoms also show the possibility of emission of high-frequency GWs \cite{gratoms}. GWs decouple almost immediately after they are created which means that, contrary to electromagnetic (EM) waves they propagate freely even before the emission of cosmic microwave background radiation. Therefore, they may be the only way to access information about these events. 
Unfortunately, the sensitivity of today's detectors does not allow for the observation of GWs coming from cosmological sources due to the very strong Big Bang Nucleosyntesis (BBN) bound \cite{Maggiore2}. BBN correctly predicts the primordial abundances of light elements such as deuterium or helium. The agreement between prediction and observation puts constraints on additional forms of energy density present at the time of BBN such as the energy density in GWs that are not included in the BBN computation. On the other hand, GWs coming from astrophysical sources - such as binaries of BHs or compact objects - do not have the same constraints since they are emitted after the BBN. The strain of GWs coming from these objects is increased as the distance separating the source from the detector decreases. Therefore primordial BHs which have not evaporated today may form binaries located nearby, whose gravitational radiation can be detected. 

To observe these high-frequency GWs, the detection mechanism we propose uses the interaction between EM waves and GWs through the graviton-to-photon conversion, otherwise known as the inverse Gertsenshtein effect. For a simple derivation of the Gertsenshtein effect, see Ref.~\cite{Gert}. The production and detection of GWs in the laboratory through this mechanism has been studied in the literature \cite{boccaletti, DeLogi}. More recent studies have looked at GW generation in the laboratory using optical methods or EM waves \cite{genoptical, genstanding, genlasers}. Detection in the laboratory using a constant magnetic field  was studied more recently in Ref.~\cite{Cavity}. Magnetic fields of cosmological size can act as detectors as well \cite{cosmodetectors}. In Ref.~\cite{upper}, bounds on the amplitude of GWs were found with existing facilities used for the detection of weakly interacting particles, such as ALPS I \cite{alps} or proposed facilities such as JURA \cite{jura}, whereas in Ref.~\cite{haloscopes} it was shown that axion haloscopes can be used as GW telescopes. In this study we focus on a new experiment for detecting GWs involving a high-energy laser as opposed to a constant magnetic field. An incoming GW interacting with the laser will be converted into an EM signal. We are interested in GWs whose frequencies are of the same order of magnitude as the laser frequencies. Since lasers can operate in a wide range of frequencies (from $10^{13}$ Hz to $10^{19}$ Hz roughly) this will allow for the search of high-frequency GWs over many orders of magnitude. The goals of this study will be to describe the interaction between GWs and the laser beam and determine the range of GW strain that are detectable for today's laser energies. 

The rest of the paper is organised as follows. In section \ref{detection}, the interaction between the laser beam and a GW in the form of a plane wave is described and we find the minimal strain of detectable GWs in terms of the laser parameters. The details for parts of the calculations are given in \ref{computation}. In section \ref{bounds1}, we apply the results found in the previous section to various lasers and find their performance. We use natural units $ \hbar = k_{B} = c =\epsilon_0 = \mu_{0} =1$ unless specified otherwise and the metric signature $(+,-,-,-)$. The Greek letters $\mu, \nu$ etc, run over spacetime indices, whereas Latin indices $i,j$ etc run over space indices.

\section{Detection}
\label{detection}
We consider a GW in the form of a monochromatic plane wave of frequency $\omega_g$ generated by an arbitrary source. It interacts with a laser beam described by a plane wave of frequency $\omega$. The two waves are assumed to be counter propagating, and we choose the coordinate system such that they are aligned along the $\hat{x}$ axis. A detector is placed along the $\hat{x}$ direction to measure the scattered EM wave (see Fig. \ref{setup}). As it will be shown below, a resonance occurs only for scattered waves of frequency $\omega_g-\omega$ (assuming $\omega_g > \omega$) when $\omega_g = 2 \omega$. Therefore, to study this resonance, we make the assumption $\omega_g > \omega$ for the rest of the paper.
\begin{figure}[H]
\centering
\includegraphics[width=1\linewidth]{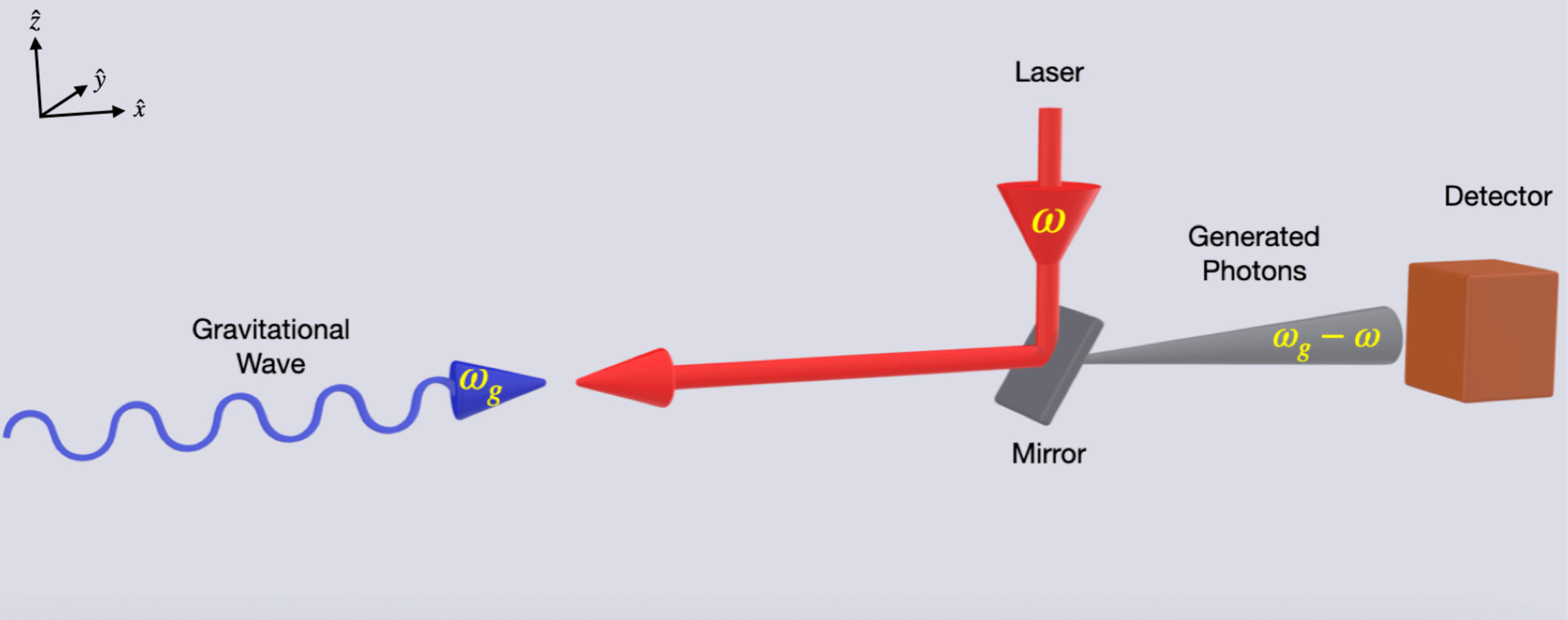}
\caption{Diagram of the experimental setup. An incident GW interacts with a linearly polarised counter-propagating laser beam. Through this interaction an EM wave is produced which does not have the same polarization as the laser beam and passes through the mirror. The distance perpendicular to the interaction region over which the laser beam propagates before being deflected by the mirror is considered small compared to the laser pulse length. The length of the interaction region is then taken to be the laser pulse length. }
\label{setup}
\end{figure}

In this section we compute the energy of the scattered wave and find bounds on the strain of the GW. We start from the Maxwell action in curved space:
\begin{equation}
    S= \int d^4 x \sqrt{- g }\left(- \frac{1}{4} g^{\mu \alpha} g^{\nu \beta} F_{\mu \nu} \; F_{\alpha \beta} \right) \;
\end{equation}
where $F_{\mu \nu}$ is the EM tensor, $g_{\mu \nu}$ is the metric tensor and $g = \text{det} (g_{\mu \nu})$. In the presence of a GW, the metric can be written as:
\begin{equation}
    g_{\mu \nu} = \eta_{\mu \nu} + h_{\mu \nu} \;,
\end{equation}
where $\eta_{\mu \nu}$ is the usual Minkowski metric and $|h_{\mu \nu}| \ll 1$. In what follows, we always keep only the leading order terms in $h_{\mu \nu}$ in the calculations. The equations of motion are given by:

\begin{equation}
 \partial_{\nu} F^{\nu \mu} = j_{\text{eff}}^{\mu} \; , 
\label{maxwell}
\end{equation}
which are the Maxwell equations in curved space and where we defined the effective four-current: 
\begin{equation}
    j_{\text{eff}}^{\mu} =  \partial_{\nu} \left(  \frac{1}{2} h F^{\mu \nu} + h^{\nu}_{\; \; \alpha} F^{\alpha \mu} - h^{\mu}_{\; \;  \alpha} F^{\alpha \nu}\right) \; .
    \label{effcurrent}
\end{equation}
Combining the Maxwell's equations in \eqref{maxwell} with Gauss' law $\Vec{\nabla} \cdot \Vec{B} = 0$ and Faraday's law $\Vec{\nabla} \times \Vec{E} + \partial_t \Vec{B} = 0$ (which are unaffected by the GW), one can write separate equations for the electric and magnetic field:
\begin{equation}
   \square \Vec{E} = -\partial_t \Vec{j}_{\rm{eff}} - \Vec{\nabla} j^0_{\rm{eff}}\;, \qquad \square \Vec{B} = \Vec{\nabla} \times \Vec{j}_{\rm{eff}} \; .
  \label{maxwell2} 
\end{equation}
The current $j_{\text{eff}}^{\mu}$ is not invariant between frames even at $\mathcal{O}(h)$. Therefore, to compute the current, one must choose a frame. A convenient choice would be the proper detector (PD) frame which is the preferred frame defined by the laboratory. The metric perturbation in this frame was found in Refs.~\cite{Marzlin,Rakhmanov}. We assume that the GW is well described by a monochromatic plane wave. The Riemann tensor is then $R_{\mu \nu \rho \sigma} \propto e^{i \omega_g(t-x)}$.  Using the fact that the Riemann tensor is invariant between frames at $\mathcal{O}(h)$, we choose to write it in the transverse traceless (TT) frame for convenience since computations are more easily done in this frame. We neglect terms of the form $a_i x^i$ and $\omega_{\text{r},i} x^i$ where $a_i$ is the acceleration due to the Earth's gravitational field and $\omega_{\text{r},i}$ the Earth's angular velocity since we consider a region of space for which $a_i x^i, \omega_{\text{r},i} x^i \ll 1$. We choose the origin of the coordinate system to be the center of the interaction region. With these approximations, the metric in the PD frame takes the form \:\footnote[1]{The result in Eqs.~\eqref{summedh} was first derived in Ref.~\cite{Cavity}.}:
\begin{subequations}\label{summedh}
\begin{align}
& h_{00}= - \omega_g^2 h_{ab}^{\text{TT}} x^a x^b \left[ -\frac{i}{\omega_g x} + \frac{1-e^{-i\omega_g x}}{(\omega_g x)^2} \right] \label{summedha}\\
& h_{ij} = \omega_g^2 \left[ (\delta_{i x} h_{j a}^{\text{TT}} + \delta_{jx} h_{i a}^{\text{TT}})x x^a - h^{\text{TT}}_{ij} x^2 - \delta_{ix} \delta_{jx} h_{a b}^{\text{TT}} x^a x^b \right] \left[  - \frac{1+e^{-i\omega_g x}}{(\omega_g x)^2} -2i\frac{1-e^{-i\omega_g x}}{(\omega_g x)^3} \right] \\
& h_{0i} = - \omega_g^2 \left[ h_{ia}^{\text{TT}} x x^a - \delta_{i x} h_{ab}^{\text{TT}} x^a x^b \right] \left[  -\frac{i}{2 \omega_g x}- \frac{e^{-i\omega_g x}}{(\omega_g x)^2} -i\frac{1-e^{-i\omega_g x}}{(\omega_g x)^3} \right] \label{summedhc}\; 
\end{align}
\end{subequations}
where $a,b = y,z$ are the components perpendicular to the direction of propagation of the GW and $h_{ij}^{\text{TT}}\propto e^{i \omega_g t}$ is the metric perturbation in the TT frame evaluated at the origin of the coordinate system~\cite{Marzlin,Rakhmanov}. The non-vanishing components of $h_{ij}^{\rm{TT}}$ are $h_{22}^{\rm{TT}} = - h_{33}^{\rm{TT}} = h_{+}$ and $h_{23}^{\rm{TT}} =  h_{32}^{\rm{TT}} = h_{\times}$, where $h_+$ and $h_{\times}$ are the plus and the cross polarizations respectively.

Due to the large energy per laser pulse and the high photon occupation number, the EM fields of the laser beam are well described classically as external fields. Furthermore, we consider them to be plane waves since the laser pulse length is much larger compared to the wavelength. In addition, we choose, for simplicity, the laser to be linearly polarised such that the third component of the electric field vanishes, $E_z = 0$. The non vanishing components of the EM tensor then are $F^{20}=-F^{21}=E_0 e^{-i \omega (t+x)}$, with $E_0$ the amplitude of the electric field. The minus sign in the exponential is chosen because we are interested in scattered EM waves whose frequency is $\omega_g-\omega$. This frequency allows for a resonance effect as it will be seen below. The scattered wave with frequency $\omega_g+\omega$ does not allow for the same resonance. Replacing these expressions as well as Eqs.~\eqref{summedh} into Eq.~\eqref{maxwell}, one obtains for the effective four-current:
\begin{equation}
    j_{\text{eff}}^{\mu} = E_0 e^{i (\omega_g - \omega)t - i \omega x} \omega_g
\begin{bmatrix}
& (h_+ y + h_{\times} z)(\omega_g f_1(u) + i \omega f_2(u)) \\
& -i(\omega - \omega_g) (h_+ y + h_{\times} z) f_2(u) \\
& h_{+} (f_2(u) + (\partial_u-i)(u^2 f_1(u))) \\
& h_{\times} (f_2(u) + (\partial_u-i)(u^2 f_1(u))) \\
\end{bmatrix} \; ,
\end{equation}
where $u = \omega_g x$ and :
\begin{equation}
f_1(u) =  - \frac{1+e^{-iu}}{u^2} -2i\frac{1-e^{-iu}}{u^3} \;, \qquad f_2(u) = -i \frac{1- e^{-i u}}{u^2} - \frac{1}{u} +\frac{i}{2} \; .
\end{equation}
The interaction region is assumed to be a rectangular cuboid with sides of length $b \times b \times L$, $b \leq L\; $, with $L$ the pulse length, satisfying $b\omega_g \sim b\omega \gg 1$. We wish to measure the fields at a distance $R$ much larger that the typical length of the interaction region. Explicitly, we assume that $\omega b^2/R \ll 1$. If $L \gg b$ we additionally assume $\omega L^4/R^3 \ll 1$. In this case, solving Eq.~\eqref{maxwell2} for one component as an illustration, we obtain (see \ref{computation}):
\begin{multline}
E_y(t,\Vec{x}) = E_0 \omega_g h_+ \frac{e^{i(\omega_g -\omega)(t-R)}}{4 \pi R}b^2 \text{sinc}\left[ \frac{b(\omega_g -\omega)n_2}{2}\right] \text{sinc}\left[ \frac{b(\omega_g -\omega)n_3}{2}\right]\\ 
\times \int_{-L \omega_g/2}^{L \omega_g/2} du \; e^{i (1 -\frac{\omega}{\omega_g})n_1 u -i \frac{\omega}{\omega_g} u} \left\{ -f_1(u) -if_2(u) -i\left( 1- \frac{\omega}{\omega_g}\right) (\partial_u - i) [u^2 f_1(u)]\right\} \; ,
\label{eyexplicit}
\end{multline}
where $n_1 = \cos{\phi} \sin{\theta}, n_2 = \sin{\phi} \sin{\theta}, n_3 = \cos{\theta}$ are the three spatial directions in spherical coordinates. Assuming $b(\omega_g - \omega) \gg 1$, the sinc functions are maximised for $n_2, n_3 \ll 1$, which is the case for our choice of detector location. Therefore, we expand around $\theta = \pi/2, \phi =0, 2 \pi $. Neglecting second order terms, $n_1 =1$. The integral in Eq.~\eqref{eyexplicit} to leading order in $L \omega_g$ is given by (for $a = \omega/ \omega_g \neq 1/2$): 
\begin{equation}
 \frac{2(1-a)\sin(a L \omega_g)}{a} + \frac{3-2a}{2a-1}\sin\left[(2a-1)\frac{L \omega_g}{2}\right] + 2 i (a-1)(- 2 i \pi \pm 2 i \pi)  \; ,
 \label{a}
\end{equation}
where the $+$ sign corresponds to $a>1/2$ and the $-$ sign corresponds to $a<1/2$. Since $a = \mathcal{O}(1)$, for $a\neq1/2$, the expression above is $\mathcal{O}(1) $. When $a \xrightarrow[]{}1/2$, the integral in Eq.~\eqref{eyexplicit} to leading order is $\mathcal{O}(L \omega_g)$, as it can be seen as well from the second term in \eqref{a}. We assume $\omega_g = 2 \omega$ for the rest of the computation to study the resonance. We give the result when $\omega_g \neq 2\omega$ below. In the case of the resonance, Eq.~\eqref{eyexplicit} is given by (along with the other components of the EM fields) Eq.~\eqref{allcomp}. The total power reaching the detector which is located at a distance $R$ from the interaction region is:

\begin{equation}
    P = \int d \Omega \; R^2 \Vec{n} \cdot \Vec{S}.
\label{power1}  
\end{equation}
where $\Vec{S} = \Vec{E} \times \Vec{B}$ is the Poynting vector. 
The integral is highly peaked around the direction $\hat{n} = (1,0,0)^{\text{T}}$. Therefore assuming our detector is large enough, we can take the integral over the whole sphere. Using Eq.~\eqref{longint}, the total power in Eq.~\eqref{power1} is then:
\begin{equation}
    P = 2 E_0^2 \cos^2\left[\omega(t-R)\right] b^2 (h_+^2 + h_{\times}^2) (2 L^2 \omega^2 + 1) \; ,
\label{totalpower}    
\end{equation}
where the first term in the last factor of Eq.~\eqref{totalpower} corresponds to the contribution of the transverse components of the EM fields and the second term is the contribution of the longitudinal component. We then compute the total energy entering the detector during a laser pulse $\tau = L$:
\begin{equation}
    E_{pp} = \int_0^L dt \; P \approx  E_0^2 b^2 L (h_+^2 + h_{\times}^2) (2 L^2 \omega^2 + 1) \; ,
\end{equation}
where we assumed $L \gg \omega^{-1}$. The laser energy per pulse is $E_{las} = E_0^2 b^2 L/2$ and since $L \omega \gg 1$, we can write:
\begin{equation}
    E_{pp} = 4 E_{las} (h_+^2 + h_{\times}^2) L^2 \omega^2 \; .
\label{energyperpulse}    
\end{equation}
For $\omega_g \neq 2 \omega$, $E_{pp} \sim E_{las}( h_{+}^2 + h_{\times}^2)$ which is smaller by a factor of $(L \omega)^2$.  The number of photons entering the detector is simply $N_{\gamma} = E_{pp} n_s/\omega$, where $n_s$ is the number of shots during the experiment.  We assume that single photon count is possible \cite{Thzsinglephoton}. Therefore, to detect a signal, at least one photon has to enter the detector: $N_{\gamma} \geq 1$. For $h_{\times} = h_+ = h$, restoring the units, the minimal strain is given by:
\begin{equation}
\begin{aligned}
&  h_{\rm{min}} = \sqrt{\frac{\hbar }{8 n_s \omega \tau^2 E_{las}}}, \qquad  \omega_g = 2 \omega \\
& h_{\rm{min}} \sim \sqrt{\frac{\hbar(\omega_g- \omega) }{2 n_s  E_{las}}}, \qquad  \omega_g \neq 2 \omega \; .\\
\end{aligned}
\label{hmin}    
\end{equation}
The results differ from the ones found in \cite{upper} where a constant magnetic field was used. The reason for this difference resides in the fact that in Ref.~\cite{upper} the TT frame was used whereas in our case the PD frame has been used. We have performed the computations in the PD frame because it corresponds to the laboratory frame and is usually the frame in which the EM fields are measured. 
\section{Projected bounds}
\label{bounds1}
\begin{figure}[H]
\centering
\includegraphics[width=0.7\linewidth]{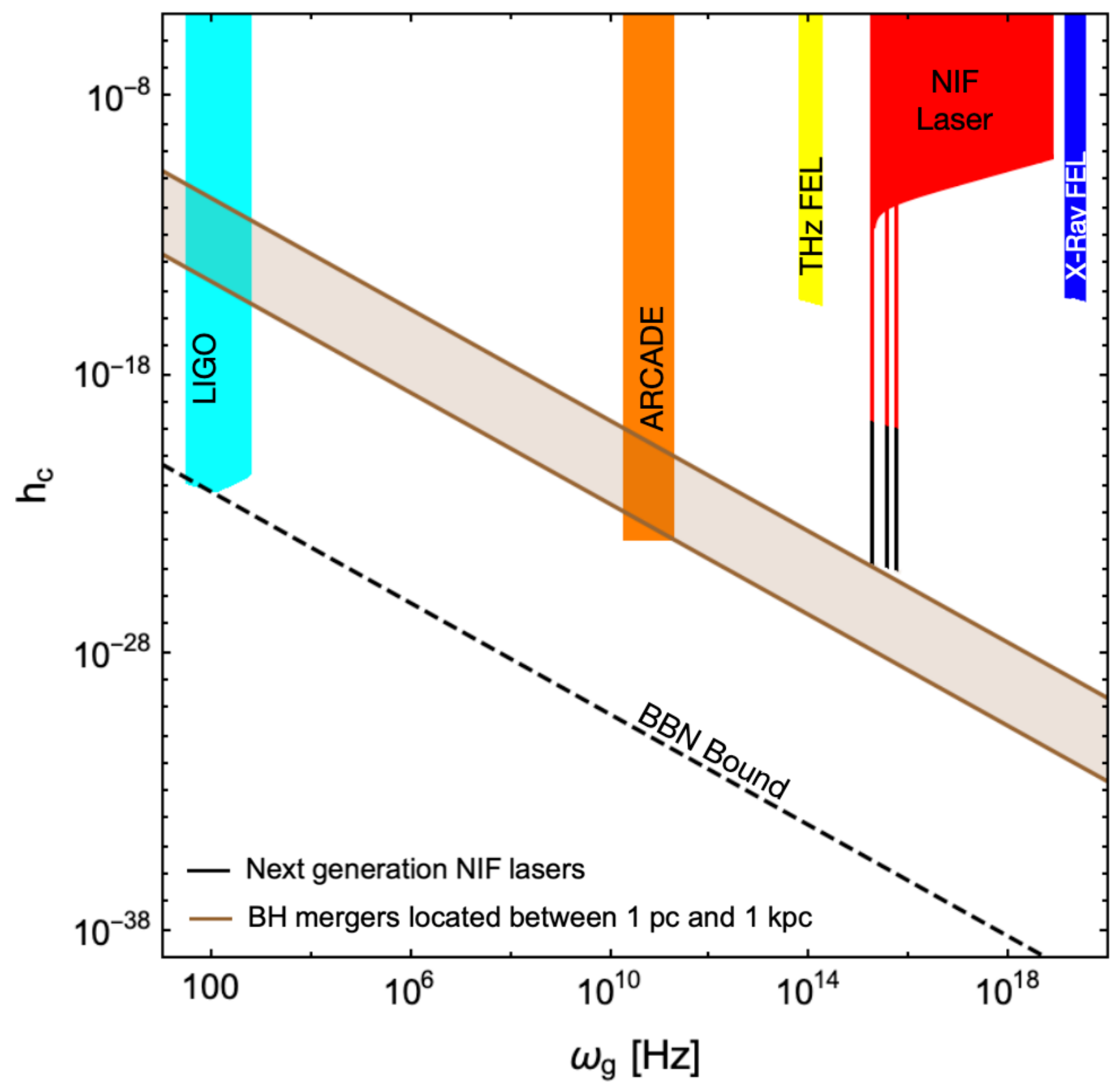}
\caption{Sensitivity of existing detectors as well as different lasers. Data from ARCADE \cite{arcade} constrain $h_c \lesssim 10^{-24}$ in the GHz range \cite{cosmodetectors}. For illustration purposes we also show the LIGO sensitivity \cite{Ligo2}. The potential sources considered are  mergers of BH binaries made of objects of equal mass located at the Innermost Stable Circular Orbit (ISCO). The ISCO radius corresponds to the minimal radial distance beyond which stable circular orbits are no longer allowed \cite{Maggiore}. All cosmological sources are located below the BBN bound and are out of reach. The characteristic strain $h_c$ for BH binaries is related to the GW amplitude $h$ by a factor of $\sqrt{2 \dot{f}/f^2}$ \cite{sqrt2fdot/f}, which is $\mathcal{O}(1)$ \cite{Maggiore}. Therefore, we use the two quantities interchangeably.  For the bounds of XFEL and the THz undulator beamline, the first case in Eq.~\eqref{hmin} was used. Since the NIF laser operates at fixed frequencies, the resonance in Eq.~\eqref{hmin} can be applied only for three frequencies depicted by the red peaks. We also show projected bounds for next generation NIF laser which are expected to have higher repetition rates.}
\label{bounds}
\end{figure}

The result \eqref{hmin} allows for the search of GWs with frequencies in the THz, optical and X-ray regimes. For illustration purposes, we consider three types of lasers, each working in a different frequency range. For each laser we assume that the duration of the experiment is one day. 
\begin{enumerate}
    \item \textbf{THz regime}: We assume we use the THz free electron laser \cite{THzregime} having the following characteristics: $\tau = 1 $ ps, $n_s = 1.7 \cdot 10^{10}$ (repetition rate of 200 kHz), 1 THz$<\omega/ 2 \pi< 30$ THz and $E_{las} = 100 \; \mu$J. 
    \item \textbf{Optical regime}: We assume we use the National Ignition Facility (NIF) laser \cite{NIF} which has a total of 192 beamlines. For each beam: $\tau = 20 $ ns, $n_s = 4$, and $E_{las} = 9.4$ kJ. The NIF laser can, in principle, operate at three different wavelengths: 1053, 527 and 351 nm (or in terms of frequency $\omega / 2 \pi = 2.4 \times 10^{14} \; \rm{Hz}, 5.7 \times 10^{14} \; \rm{Hz}, 8.5 \times 10^{14} \; \rm{Hz}$).
    \item \textbf{X-ray regime}: We assume we again use a free-electron laser, but operating in the X-ray regime such as the European XFEL or the one at the SLAC National Accelerator Laboratory \cite{XrayFEL}. The characteristics are $\tau = 0.1$ ps, $n_s = 8.6 \times 10^5$ (repetition rate of 10 Hz), and $5.8$ keV $<\omega/ 2\pi < 24$ keV. The energy per pulse depends on the frequency of the beam. Between 5.8 and 9.3 keV, $E_{las}= 2$ mJ. Between 9.3 and 12 keV, $E_{las}= 1$ mJ and between 12 and 24 keV $E_{las}= 0.5$ mJ.
\end{enumerate}
The bounds on the strain of detectable GWs for the different lasers are shown in Fig.~\ref{bounds}. 

As the NIF laser operates at three fixed frequencies, the resonance condition $\omega_g = 2 \omega$ is valid for three GW frequencies. This can be seen by the three red peaks in Fig.~\ref{bounds}. For other frequencies, the non resonant result has to be taken. On the other hand, the XFEL and THz undulator beamline operate in continuous frequency ranges. Therefore the resonant result in Eq.~\eqref{hmin} can be taken throughout the operating frequency range. 
The bounds for these lasers are:
\begin{subequations}
\label{3bounds}
\begin{align}
& h^{\rm{THz}}_{\rm{min}} = 2.3 \times 10^{-16} \left( \frac{1.7 \times10^{10}}{n_s}\right)^{1/2} \left( \frac{ 30 \; \rm{THz}}{\omega_g/2 \pi}\right)^{1/2} \left( \frac{1 \; \rm{ps}}{\tau}\right) \left( \frac{100 \; \mu\rm{J}}{E_{las}}\right)^{1/2}\\
&  h^{\rm{opt}}_{\rm{min}} = 1.8 \times 10^{-20} \left( \frac{4}{n_s}\right)^{1/2} \left( \frac{8.5 \times 10^{14} \; \rm{Hz}}{\omega_g /2 \pi}\right)^{1/2} \left( \frac{20 \; \rm{ns}}{\tau}\right) \left( \frac{9.4 \; \rm{kJ}}{E_{las}}\right)^{1/2}\\
&  h^{\rm{Xray}}_{\rm{min}} = 1.3 \times 10^{-16} \left( \frac{8.6 \times10^{5}}{n_s}\right)^{1/2} \left( \frac{1.4 \times10^{19} \; \rm{Hz}}{\omega_g}\right)^{1/2} \left( \frac{0.1 \; \rm{ps}}{\tau}\right) \left( \frac{2 \; \rm{mJ}}{E_{las}}\right)^{1/2} \; .
\end{align}
\end{subequations}
In the optical regime, we can also produce projected bounds for a new generation laser. For this new NIF laser, we assume that all 192 beams are available with a total energy $E_{las} = 1.8 \; $MJ, and a repetition rate of $\sim 10$ kHz and the duration of the experiment being one day. In this case the lower bound becomes: 
\begin{equation}
    h^{\rm{opt,fut}}_{\rm{min}} = 6.3 \times 10^{-26}  \left( \frac{8.6\times10^{8}}{n_s}\right)^{1/2} \left( \frac{ 8.5 \times 10^{14} \rm{Hz}}{\omega_g/2 \pi}\right)^{1/2} \left( \frac{20 \; \rm{ns}}{\tau}\right) \left( \frac{1.8\;\rm{MJ}}{E_{las}}\right)^{1/2} \; .
\label{futurebounds}    
\end{equation}
For the X-Ray regime instead, we can augment the sensitivity by increasing the interaction time between the GW and the laser beam. Suppose that the laser beam enters a region separated by two crystal planes. The beam is reflected every time it reaches a crystal plane. If during that time the beam interacts with a GW,  a scattered EM wave is produced. At the resonant frequency, the scattered EM wave has the same frequency as the incoming one and is propagating in the opposite direction being reflected every time it reaches a crystal plane. The total energy reaching the detector found in  Eq.\eqref{energyperpulse} is then enhanced by the number of times the beam is reflected. This number is given by $n_{\rm{ref}} = l /(c \tau \cos{\theta_B})$, where $l$ is the length of the crystal and $\theta_B$ the Bragg angle. If the distance separating the two crystals is $c \tau \sin{\theta_B}$, with $l \sim 1$ m, $n_{\rm{ref}} \sim 10^{4}$, we find for the lowest bound:
\begin{equation}
h^{\rm{Xray}}_{\rm{min}} = 1.3 \times 10^{-18} \left( \frac{8.6 \times10^{5}}{n_s}\right)^{1/2} \left( \frac{1.4 \times10^{19} \; \rm{Hz}}{\omega_g}\right)^{1/2} \left( \frac{0.1\; \rm{ps}}{\tau}\right) \left( \frac{2 \; \rm{mJ}}{E_{las}}\right)^{1/2} \left( \frac{10^4 }{n_{\rm{ref}}}\right)^{1/2} \;.
\end{equation}

As stated in the introduction, GWs created in the early Universe by cosmological sources are severely bounded from the BBN bound. BBN predicts successfully primordial abundances of light elements without taking into account the energy density in GWs. Therefore any additional energy density considered must be small enough to not spoil the BBN results. The constraint takes the form \cite{Challenges, bbn1}:
\begin{equation}
   \Omega_{GW} = \frac{4 \pi^2}{3 H_0^2} \left( \frac{\omega_g}{ 2\pi}\right)^2h_c^2\lesssim 3 \times10^{-6}, \qquad h_c \lesssim 10^{-33} \left( \frac{10^{12} \rm{Hz}}{\omega_g / 2 \pi}\right)
\end{equation}
where $H_0$ is the Hubble rate of expansion today. $\Omega_{GW} = \rho_c^{-1} (d \rho_{GW}/ d \ln{f})$ represents the energy density of GWs $\rho_{GW}$ per logarithmic frequency normalised to the critical energy density today $\rho_c$.  These sensitivities are out of reach for today's laser detectors. The reason for this is due to the brief interaction duration imposed by the pulse of the laser. GWs from cosmological sources form a stochastic background which can interact continuously with a detector. Therefore, for these sources it is convenient to consider a constant magnetic field instead of a laser as it has been done in \cite{upper}. On the other hand, coherent sources such as binary BH mergers can be detected if they are close to the detector since their strain decreases as $h \propto D^{-1}$, with $D$ the distance to the source \cite{Maggiore, postnewtonian}. The highest frequency GWs for a given binary are emitted just before coalescence, when the objects of the binary are located on the Innermost Stable Circular Orbit (ISCO). The ISCO frequency scales as $f_{\rm{ISCO}} \propto m^{-1}$, where $m$ is the total mass of the binary \cite{Maggiore}. Detecting high-frequency GWs can therefore lead to the discovery of light binaries in the nearby interstellar space.

In Eqs.~\eqref{3bounds}, \eqref{futurebounds}, it is assumed that the frequency of the GW is the resonant frequency throughout the duration of the experiment, which is one day. But for BH binaries, the frequency changes rapidly close to the coalescence. We therefore wish to find the duration for which the frequency of the GW emitted by the binary satisfies the resonance condition. We first assume that the GW frequency is not exactly twice the laser frequency but that it is $\omega_g+ \delta \omega_g $, with $\omega_g = 2\omega$. From Eq.~\eqref{a}, we find that for the EM fields to be enhanced by a factor of $(L \omega)$:
\begin{equation}
    \frac{\delta \omega_g}{\omega_g} \ll \frac{1}{ L \omega} \ll 1 \; .
   \label{deltaomega} 
\end{equation}
The frequency of a GW emitted by a BH binary is \cite{Maggiore}:
\begin{equation}
    \omega_{gw}(\Tilde{t}) = 2 \left( \frac{5}{256 \Tilde{t} }\right)^{3/8} (G M_c)^{-5/8}, \quad \Tilde{t} = t_{\rm{coal}}-t
\label{freqvaria}    
\end{equation}
where  $M_c$ is the chirp mass and $t_{\rm{coal}}$ is the time of the coalescence. We assume that the GW frequency changes from $2 \omega$ to  $2 \omega + \delta \omega_g$ in a duration $\Delta t$. Then using Eqs.~\eqref{deltaomega} and \eqref{freqvaria}, we find that for the resonance condition to still be applicable the duration must satisfy:
\begin{equation}
    \Delta t \ll \frac{5 c^6}{24 (G m)^{5/3} L \omega^{11/3}} \sim 10 \;  \text{days}  \left( \frac{1 \; \rm{ps}}{\tau}\right) \left( \frac{30 \; \rm{THz}}{2 \omega/ 2 \pi}\right)^{11/3} \left( \frac{10^{-22} \; M_{\odot}}{m}\right)^{5/3}
\end{equation}
where we restored the units and with $m$ the total mass of the binary assuming an equal mass binary. We have replaced the laser parameters by the ones for a THz laser as an illustration. We find that when the total mass of the binary is $m \sim  10^{-22} M_{\odot}$, GWs are emitted at the resonant frequency  for a duration comparable to the experiment duration. Therefore, the heavier the binary, the shorter the duration of emission of resonant GWs. There are two ways for the frequency emitted to be resonant throughout the experiment. Either the total mass of the binary is small enough or there are multiple heavier binaries emitting at the resonant frequency such that at any moment during the experiment there is at least one binary emitting GWs at the desired frequency. For both cases, the sensitivity is the one found in Eqs.~\eqref{3bounds}, \eqref{futurebounds} - but in the first case the astrophysical reach is reduced due to the smaller mass. Nonetheless, lighter binaries are still of interest. Assuming dark matter is made of primordial BHs, the characteristic size of a region containing at least a merger per unit time decreases as the mass of the merger decreases \cite{hunt}. Therefore, even though the astrophysical reach is smaller for light binaries, their density can be higher. 
\section{Conclusion}
We studied the interaction between a planar GW and a high-energy laser beam. By measuring the energy of the scattered EM wave, we have obtained lower bounds on the strain of the GW that can be detected by laser beams. It was found that a resonance occurs when the frequency of the GW is twice the frequency of the laser beam. In this particular case, the energy of the scattered EM wave was enhanced by a factor of $(\omega \tau)^2$. Applying the general result found for existing lasers, we find lower bounds in three different frequency ranges: the THz, the optical and the X-ray ranges. Although the minimal strain decreases as $h_{\rm{min}}\propto \omega^{-1/2}$, it was found that the most sensitive laser detector operates in the optical range (and not in the X-ray) because of the high-energy and long pulses available for optical lasers. In this case, $h_{\rm{min}} \sim 10^{-20}$. We computed sensitivities for future laser facilities as well. By assuming a higher repetition rate (10 kHz) for the NIF laser, the minimal strain is then given by $h^{\rm{opt}}_{\rm{min}} \sim 10^{-26}$. The limitation of this detection mechanism is that it allows for the detection of a monochromatic GW, the one for which $\omega_g = 2 \omega$. GWs generated by cosmological sources cannot be detected due to their weak strain imposed by the BBN bound. Regarding astrophysical sources, only mergers made of very light BHs or exotic compact objects \cite{compact1, compact2} can emit GWs in the frequency range of interest. We looked at the former source because the GW strain is higher in this case. For these sources, the GW frequency increases as time passes and the duration during which the emitted GWs have a resonant frequency is short. This duration is increased either for smaller binary masses or if we consider multiple heavier binaries with the GWs interacting continuously with the laser. In the second case the astrophysical reach is higher because of the heavier mass. We conclude by noting that the detection of GWs at high frequency remains relatively unexplored. Even with current laser facilities, the proposed laser experiments have the potential of unlocking novel physics.     
\section{Acknowledgements }
We thank Prof Pedro Ferreira and Dr James Marsden for the helpful discussions and comments on this manuscript. 
\appendix
\section{Computation}
\label{computation}
Writing explicitly Eq.~\eqref{maxwell2} gives: 
\begin{equation}
\begin{aligned}
& \square E_x = E_0 e^{i (\omega_g - \omega)t - i \omega x} \omega_g^3 (h_+ y + h_{\times} z) \left[ 2 i \frac{\omega}{\omega_g} f_1(u) + \left( 1- \frac{2 \omega}{\omega_g}\right)f_2(u) - f_1'(u)\right] \\ 
& \square E_y = E_0 e^{i (\omega_g - \omega)t - i \omega x} \omega_g^2 h_+ \left[ -f_1(u) -i f_2(u) -i\left( 1- \frac{\omega}{\omega_g}\right) (\partial_u - i) (u^2 f_1(u))\right] \\ & \square E_z = E_0 e^{i (\omega_g - \omega)t - i \omega x} \omega_g^2 h_{\times} \left[ -f_1(u) -if_2(u) -i\left( 1- \frac{\omega}{\omega_g}\right) (\partial_u - i) (u^2 f_1(u))\right] \\
& \square B_x = 0 \\
& \square B_y = E_0 e^{i (\omega_g - \omega)t - i \omega x} \omega_g^2 h_{\times} \left[ i f_2(u) + \left( \frac{i \omega}{\omega_g} -\partial_u\right) (\partial_u -i)(u^2 f_1(u)) + f_1(u) \right] \\ 
& \square B_z = -E_0 e^{i (\omega_g - \omega)t - i \omega x} \omega_g^2 h_{+} \left[ i f_2(u) + \left( \frac{i \omega}{\omega_g} -\partial_u\right) (\partial_u -i)(u^2 f_1(u)) + f_1(u) \right] 
\end{aligned} 
\end{equation}
where $u =\omega_g x$. The solutions for the generated EM fields are of the form:
\begin{equation}
    E_y(t,\Vec{x}) = \int d^3 y \;  e^{i(\omega_g - \omega)(t - |\Vec{x} - \Vec{y}|) - i \omega y_1} \frac{f_{E_2}(\Vec{y})}{4 \pi |\Vec{x} -\Vec{y}|} \; ,
\end{equation}
where the integral is taken over the volume of the interaction region, which we assume to be a rectangular cuboid with sides of length $b \times b \times L$, $b \leq L,\; b\omega_g \gg 1$. For $|x|\gg |y|$, the denominator is just $4 \pi |x|$. Regarding the exponent, assuming that $\omega_g -\omega \sim \omega$, we Taylor expand:
\begin{equation}
    \omega |\Vec{x}-\Vec{y}| = \omega R -\omega \Vec{n}\cdot\Vec{y}+ \frac{\omega y^2}{2 R} - \frac{\omega (\Vec{n}\cdot\Vec{y})^2}{2 R} + \frac{\omega y^2 \Vec{n}\cdot\Vec{y}}{2 R^2} - \frac{\omega  (\Vec{n}\cdot\Vec{y})^3}{2 R^2} + \mathcal{O}\left(\frac{\omega y^4}{R^3}\right), \quad y\leq L \; ,
\end{equation}
where $R = |\Vec{x}|$ is the distance from the interaction region to the detector and $\Vec{n} = \Vec{x}/R \approx (1,0,0)^{\rm T}$. In the case $L\sim b$, we choose $x$ such that $\omega b^2/R \ll 1$, such that we keep only the two first terms in the expansion. In the case $L\gg b$, we additionally assume $\omega L^4/R^3 \ll 1$ which allows us to neglect the additional terms beyond this expansion. For $n_1 \approx 1$ and for $\omega/\omega_g =1/2$ (the resonance), it is straightforward to show that all the components take the form:
\begin{equation}
\begin{aligned}
& E_x = E_0  \frac{e^{i\omega(t-R)}}{4 \pi R} b^3 \omega_g^2 \frac{1}{L \omega_g} \left( h_+ \text{sinc}\left( \frac{b\omega n_3}{2}\right) \frac{\cos\left( \frac{b\omega n_2}{2}\right) - \text{sinc}\left( \frac{b\omega n_2}{2}\right) }{\frac{b\omega n_2}{2}} \right. \\ 
& \left. + h_{\times} \text{sinc}\left( \frac{b\omega n_2}{2}\right) \frac{\cos\left( \frac{b\omega n_3}{2}\right) - \text{sinc}\left( \frac{b\omega n_3}{2}\right) }{\frac{b\omega n_3}{2}}\right) \\
& E_y = E_0  \frac{e^{i\omega(t-R)}}{4 \pi R} \omega_g^2 h_+ L b^2 \text{sinc}\left( \frac{b\omega n_2}{2}\right) \text{sinc}\left( \frac{b\omega n_3}{2}\right) \\
& E_z =  E_0  \frac{e^{i\omega(t-R)}}{4 \pi R} \omega_g^2 h_{\times} L b^2 \text{sinc}\left( \frac{b\omega n_2}{2}\right) \text{sinc}\left( \frac{b\omega n_3}{2}\right) \\
& B_x = 0 \\
& B_y =  -E_0  \frac{e^{i\omega(t-R)}}{4 \pi R} \omega_g^2 h_{\times} L b^2 \text{sinc}\left( \frac{b\omega n_2}{2}\right) \text{sinc}\left( \frac{b\omega n_3}{2}\right) \\
& B_z =  E_0  \frac{e^{i\omega(t-R)}}{4 \pi R} \omega_g^2 h_{+} L b^2 \text{sinc}\left( \frac{b\omega n_2}{2}\right) \text{sinc}\left( \frac{b\omega n_3}{2}\right) \; .\\
\end{aligned}
\label{allcomp}
\end{equation}
Having all the components of the EM fields, we can compute the integral in Eq.~\eqref{power1}, to leading order in $(b \omega)^{-1}$. The integral is peaked along the direction $\hat{x}$. Therefore, assuming our detector is large enough, we can take the integral over the whole sphere. Taking the real part in the expressions above and using:
\begin{equation}
\int d \Omega \; \sin \theta \cos \phi \; \text{sinc}^2\left( \frac{b \omega}{2}\cos \theta\right) \text{sinc}^2\left( \frac{b \omega}{2}\cos \phi \sin \theta \right) \approx \frac{4 \pi^2}{(b \omega)^2}  \; ,
\label{longint}
\end{equation}
to leading order in $(b \omega)^{-1}$, we find that the total power is given by Eq.~\eqref{totalpower}.

\end{document}